\documentclass[12pt]{article}

\usepackage{amsfonts,amssymb,amsmath}
\usepackage{graphicx}
\DeclareGraphicsExtensions{epsfig}
\begin{document}
\title{\bf An improved bound for quantum speed limit time in open quantum systems by introducing an alternative fidelity }
\author{ Abbas Ektesabi  $^{b}$
\thanks{E-mail:ektesabi@azaruniv.edu}  ,
Naghi Behzadi $^{a}$
\thanks{E-mail:n.behzadi@tabrizu.ac.ir}  ,
Esfandyar Faizi $^{b}$
\thanks{E-mail:efaizi@azaruniv.edu}
\\ $^a${\small Research Institute for Fundamental Sciences, University of Tabriz, Iran,}
\\ $^b${\small Physics Department, Azarbaijan shahid madani university, Iran.}}\maketitle

\begin{abstract}
In this paper, we introduce a new alternative quantum fidelity for quantum states which perfectly satisfies all Jozsa's axioms and is zero for orthogonal states. By employing this fidelity, we derive an improved bound for quantum speed limit time in open quantum systems in which the initial states can be chosen as either pure or mixed. This bound leads to the well-known Mandelstamm-Tamm type bound for nonunitary dynamics in the case of initial pure states. However, in the case of initial mixed states, the bound provided by the introduced fidelity is tighter and sharper than the obtained bounds in the previous works.
\\
\\
{\bf PACS Nos:}
\end{abstract}

\section{Introduction}
Quantum speed limits (QSLs) are the ultimate bounds imposed by quantum mechanics on the minimal evolution time for a quantum state to become orthogonal to itself. QSLs have been widely investigated since the appearance of first major result by Mandelstamm and Tamm $\cite{Mandelstam}$. They derived a QSL limit time bound for a quantum system that evolves between two pure orthogonal initial and final states under the time independent Hamiltonian $H$. The bound is given by $\tau\geq\pi\hbar/(2\Delta E$), where $\Delta E$ is the variance of the energy. Later Margolus and Levitin $\cite{Margolus}$ provided a different QSL time bound for a closed system reads as $\tau\geq\pi\hbar/(2E)$, where $E$ is the mean energy with respect to the ground state. Both the Mandelstamm-Tamm and Margolus-Levitin bounds are attainable in closed quantum systems for initial pure states, while for general mixed states they can be rather loose. Since any system is coupled to an environment, an analogous bound for open quantum system is highly desirable. Taddei $et$ $al$. $\cite{Taddei}$ extended the Mandelstamm-Tamm type bound to both unitary and nonunitary processes described by positive non-unitary maps by using of quantum Fisher information for time estimation. However, for the case of the initial mixed states, it is hard to evaluate the bound due to minimization of the quantum Fisher information in the enlarged system-environment space. Later Deffner and Lutz $\cite{Deffner}$ extended both Mandelstamm-Tamm and Margolus-Levitin bounds to open quantum system by exploiting Caushy-schwarz and von Neumann trace inequality, respectively. They showed that the non-Markovian effect leads to the faster quantum evolution. However, their bound is derived from pure initial states and can not be applied into the mixed initial states. Also del Campo $et$ $al$. $\cite{plenio}$ derived an analytical and computable QSL bound for open quantum systems by exploiting the relative purity. Relative purity can make a distinction between two initial pure states, however it  may fail as a distance measure between two initial mixed states. Recently Sun $et$ $al$. $\cite{Zhe Sun}$ derived another quantum speed limit bound for open quantum systems by employing an alternative fidelity introduced in $\cite{Wang}$ which the initial states can be chosen as either pure or mixed. However their bound is not tight and the alternative fidelity which they used as a distance measure, fails to satisfy one of the Jozsa's four axioms $\cite{Jozsa}$.

In this paper, we first propose a new alternative definition of quantum fidelity between quantum states which perfectly satisfies all Jozsa's four axioms. Also this fidelity is zero when two density matrices are orthogonal, the criterion which can not be satisfied by some previously introduced fidelities $\cite{Miszczak,Mendonca,Chen}$. By employing this fidelity and applying Caushy-schwarz inequality, we derive a QSL time bound for open quantum systems which the initial state can be chosen as either pure or mixed. This bound leads to the Mandelstamm-Tamm type bound for nonunitary dynamics in the case of initial pure states. However, in the case of initial mixed states, the obtained bound is tighter and sharper than the bounds provided by the previous works.

The work is organized as follows. In Sec. II we introduce an alternative fidelity and discuss its basic properties. In Sec. III we derive the QSL time bound by exploiting the new fidelity. Sec. IV is devoted to demonstrate the performance of QSL time bound obtained by the intoduced fidelity, by considering a two-level atomic system coupled resonantly to a leaky vacuum reservoir. Finally, the paper is ended by a brief conclusion.

\section{Properties of the alternative fidelity}
 In order to derive a QSL time bound for open quantum systems, we should use a distance measure between two quantum states. Among the distance measures, the Bures fidelity is the most important one for quantum computation and quantum information processing $\cite{Bures,uhlmann,Nielsen}$. This fidelity for two general mixed states $\rho$ and $\sigma$ is given by

\begin{eqnarray}
F(\rho,\sigma)=\biggr(
                       Tr(\sqrt{\rho^{1/2}\sigma\rho^{1/2}})
                                            \biggr)
^2.
\end{eqnarray}
The Bures fidelity is supported widely by number of desired properties and also satisfies all Jozsa's four axioms:

    $(\mathrm{A}_{1})$  $0\leq F(\rho,\sigma)\leq1$ and $F(\rho,\sigma)=1$ if and only if $\rho=\sigma$.

    $(\mathrm{A}_{2})$  $F$ is symmetry under swapping of two states, i.e. $F(\rho,\sigma)=F(\sigma,\rho)$.

    $(\mathrm{A}_{3})$  $F(\rho,\sigma)$ is invariant under unitary transformations on the state space.

    $(\mathrm{A}_{4})$  If one of the state is pure
    ($\sigma=|\psi\rangle\langle\psi|$), the fidelity reduces to $F(\rho,|\psi\rangle\langle\psi|)=\langle\psi|\rho|\psi\rangle$.

However, due to the difficulty in calculation of Bures fidelity, there have been some attempts to find an alternative fidelity to avoid this difficulty.
Wang $et$ $al$. $\cite{Wang}$ proposed an alternative fidelity in terms of their Hilbert-Schmidt inner product and their purities, which reads as
\begin{eqnarray}
F_{1}(\rho,\sigma)=\frac{Tr(\rho \sigma)}{\sqrt{{Tr(\rho^2)}{Tr(\sigma^2)}}}.
\end{eqnarray}
Recently Sun $et$ $al$. $\cite{Zhe Sun}$ derived an analytical and computable quantum speed limit bound for open quantum systems by exploiting $F_{1}$ in Eqs. (2). However, one can easily find that $F_{1}$ fails to satisfy the 4$th$ axiom and thus it may induce some defects into derivation of quantum speed limit (as will be seen in the last section of this paper). Another fidelity was defined by Miszczak $et$ $al$. $\cite{Miszczak}$ and Mendon\c{c}a $et$ $al$. $\cite{Mendonca}$, which reads as

\begin{eqnarray}
F_{2}(\rho,\sigma)=Tr(\rho \sigma)+\sqrt{1-Tr(\rho^2)}\sqrt{1-Tr(\sigma^2)}.
\end{eqnarray}
Also Chen $et$ $al$. $\cite{Chen}$ defined the following fidelity which is essentially the same as $F_{2}$
\begin{eqnarray}
F_{3}(\rho,\sigma)=\frac{1-r}{2}+\frac{1+r}{2}F_{2}(\rho,\sigma),
\end{eqnarray}
where $r=1/(d-1)$ and $d$ is the dimension of the Hilbert space. $F_{2}$ and $F_{3}$ satisfy all Jozsa's four axioms. Although $F_{2}$ and $F_{3}$ are identical in $d=2$ and they reduce to equivalent Bures fidelity (1), However, for two orthogonal density matrices such as
\begin{eqnarray}
\begin{array}{c}
  \rho=\frac{1}{2}(|0\rangle\langle0|+|1\rangle\langle1|), \\\\
  \sigma=\frac{1}{2}(|2\rangle\langle2|+|3\rangle\langle3|),
\end{array}
\end{eqnarray}
defined on a 4-dimensional Hilbert space spanned by $\{|n\rangle, n=0, 1, 2, 3\}$, $F_{2}(\rho, \sigma)$ and $F_{3}(\rho, \sigma)$ are failed to be zero for $d>2$ (in this case $d=4$) $\cite{Wang}$.

In this paper, we give a new alternative definition of quantum fidelity between quantum states, which reads as
\begin{eqnarray}
\mathcal{F}(\rho,\sigma)=\biggr(1+\sqrt{\frac{1-Tr(\rho^2)}{Tr(\rho^2)}}\sqrt{\frac{1-Tr(\sigma^2)}{Tr(\sigma^2)}}\biggr)Tr(\rho \sigma).
\end{eqnarray}
$\mathcal{F}$ satisfies all Jozsa's four axioms and it is zero when two density matrices are orthogonal. It is not difficult to see that $\mathcal{F}$ satisfies Jozsa's axioms $(\mathrm{A}_{2})$, $(\mathrm{A}_{3})$ and $(\mathrm{A}_{4})$. In the following, we prove that $\mathcal{F}$ satisfies the axiom $(\mathrm{A}_{1})$.

Proof of axiom $(\mathrm{A}_{1})$. We rewrite the Eq. (6) as follows
\begin{eqnarray}
\mathcal{F}(\rho,\sigma)=Tr(\rho \sigma)+\sqrt{\frac{1-Tr(\rho^2)}{Tr(\rho^2)}}\sqrt{\frac{1-Tr(\sigma^2)}{Tr(\sigma^2)}}Tr(\rho \sigma).
\end{eqnarray}
By using the Cauchy-Schwarz inequality, i.e, $|Tr(\rho\sigma)|\leq \sqrt{Tr(\rho^2)Tr(\sigma^2)}$ in the second term of Eq. (7), we get
\begin{eqnarray}
\begin{split}
\mathcal{F}(\rho,\sigma)=Tr(\rho \sigma)+\sqrt{\frac{1-Tr(\rho^2)}{Tr(\rho^2)}}\sqrt{\frac{1-Tr(\sigma^2)}{Tr(\sigma^2)}}Tr(\rho \sigma)\\\leq Tr(\rho \sigma)+\sqrt{1-Tr(\rho^2)}\sqrt{1-Tr(\sigma^2)}.
\end{split}
\end{eqnarray}
Now by considering the inequality $Tr(\rho \sigma)+\sqrt{1-Tr(\rho^2)}\sqrt{1-Tr(\sigma^2)}\leq1 $  $\cite{Mendonca}$, we reach
\begin{eqnarray}
\mathcal{F}(\rho,\sigma)=\biggr(1+\sqrt{\frac{1-Tr(\rho^2)}{Tr(\rho^2)}}\sqrt{\frac{1-Tr(\sigma^2)}{Tr(\sigma^2)}}\biggr)Tr(\rho \sigma)\leq 1.
\end{eqnarray}
\hfill $\square$

The fidelity $\mathcal{F}$ is super-multiplicative under tensor products, i.e.,
\begin{eqnarray}
\mathcal{F}(\rho_{1}\otimes\rho_{2},\sigma_{1}\otimes\sigma_{2})\geq\mathcal{F}(\rho_{1},\sigma_{1})\mathcal{F}(\rho_{2},\sigma_{2}).
\end{eqnarray}
To prove this property we can write
\begin{eqnarray}
\mathcal{F}(\rho_{1}\otimes\rho_{2},\sigma_{1}\otimes\sigma_{2})=\biggr(1+\sqrt{\frac{1-Tr(\rho_{1}^2)Tr(\rho_{2}^2)}{Tr(\rho_{1}^2)Tr(\rho_{2}^2)}}\sqrt{\frac{1-Tr(\sigma_{1}^2)Tr(\sigma_{2}^2)}{Tr(\sigma_{1}^2)Tr(\sigma_{2}^2)}}\biggr)Tr(\rho_{1}\sigma_{1})Tr(\rho_{2}\sigma_{2})
\end{eqnarray}
and
\begin{eqnarray}
\begin{split}
\mathcal{F}(\rho_{1},\sigma_{1})\mathcal{F}(\rho_{2}\sigma_{2})=\biggr(1+\sqrt{\frac{(1-Tr(\rho_{1}^2))(1-Tr(\sigma_{1}^2))}{Tr(\rho_{1}^2)Tr(\sigma_{1}^2)}}\biggr)\times
\\
\biggr(1+\sqrt{\frac{(1-Tr(\rho_{2}^2))(1-Tr(\sigma_{2}^2))}{Tr(\rho_{2}^2)Tr(\sigma_{2}^2)}}\biggr)Tr(\rho_{1}\sigma_{1})Tr(\rho_{2}\sigma_{2})
\end{split}.
\end{eqnarray}
By defining $r_{i}:=Tr(\rho_{i}^2)$ and $s_{i}:=Tr(\sigma_{i}^2)$, we have to show that
\begin{eqnarray}
\begin{split}
\sqrt{(1-r_{1}r_{2})(1-s_{1}s_{2})}\geq\sqrt{(1-r_{1})(1-s_{1})}\sqrt{r_{2}s_{2}}+\sqrt{(1-r_{2})(1-s_{2})}\sqrt{r_{1}s_{1}}+\\
\sqrt{(1-r_{1})(1-s_{1})(1-r_{2})(1-s_{2})}.
\end{split}
\end{eqnarray}
To this aim, we define two vectors
\begin{eqnarray}
X=\begin{pmatrix}
\sqrt{r_{1}}\sqrt{1-r_{2}}\\
\sqrt{r_{2}}\sqrt{1-r_{1}}\\
\sqrt{1-r_{1}}\sqrt{1-r_{2}}\\
\end{pmatrix}
\qquad\text{and}\qquad
Y=\begin{pmatrix}
\sqrt{s_{1}}\sqrt{1-s_{2}}\\
\sqrt{s_{2}}\sqrt{1-s_{1}}\\
\sqrt{1-s_{1}}\sqrt{1-s_{2}}\\
\end{pmatrix},
\end{eqnarray}
with
\begin{eqnarray}
\begin{split}
\langle X|Y\rangle=\sqrt{(1-r_{1})(1-s_{1})}\sqrt{r_{2}s_{2}}+\sqrt{(1-r_{2})(1-s_{2})}\sqrt{r_{1}s_{1}}+\\
\sqrt{(1-r_{1})(1-s_{1})(1-r_{2})(1-s_{2})},
\end{split}
\end{eqnarray}
and
\begin{eqnarray}
\langle X|X\rangle=(1-r_{1}r_{2})
\qquad\text{and}\qquad
\langle Y|Y\rangle=(1-s_{1}s_{2}).
\end{eqnarray}
Now, by using the Cauchy-Schwarz inequality
\begin{eqnarray}
\sqrt{\langle X|X\rangle\langle Y|Y\rangle}\geq\langle X|Y\rangle,
\end{eqnarray}
the inequality (13) is satisfied.
\hfill $\square$

The fact that $\mathcal{F}$ is super-multiplicative and not multiplicative, may be a sign that it may not have the monotonicity property of a fidelity. However the preliminary numerical search favors the validity of monotonicity property of $\mathcal{F}$ and shows that it is monotonically increasing under completely positive trace preserving (CPTP) maps. For example, the counterexample used to show that $F_{2}$ in Eq. (3) does not behave monotonically under CPTP maps in $\cite{Mendonca}$, satisfies the desired monotonicity property of $\mathcal{F}$.
By denoting $\varrho$ and $\varsigma$ as the two two-qubit density matrices
\begin{eqnarray}
\varrho=\frac{1}{2}\begin{pmatrix}
1&0&0&0\\
0&1&0&0\\
0&0&0&0\\
0&0&0&0\\
\end{pmatrix},
\qquad\text{and}\qquad
\varsigma=\frac{1}{2}\begin{pmatrix}
0&0&0&0\\
0&0&0&0\\
0&0&1&0\\
0&0&0&1\\
\end{pmatrix},
\end{eqnarray}
and by considering the quantum operations of tracing over the first or second qubit, we have
\begin{eqnarray}
\mathcal{F}(Tr_{1}(\varrho),Tr_{1}(\varsigma))=1>0=\mathcal{F}(\varrho,\varsigma),
\end{eqnarray}
and
\begin{eqnarray}
\mathcal{F}(Tr_{2}(\varrho),Tr_{2}(\varsigma))=0=\mathcal{F}(\varrho,\varsigma).
\end{eqnarray}
Therefore, Eq. (19) and (20) show that $\mathcal{F}$ is monotonically increasing and satisfies the desired monotonicity property under this map.

Also our preliminary numerical calculations show that $\mathcal{F}$ satisfies the property of concavity, so the inequality
\begin{eqnarray}
\mathcal{F}(\rho,p\sigma_{1}+(1-p)\sigma_{2})\geq p\mathcal{F}(\rho,\sigma_{1})+(1-p)\mathcal{F}(\rho,\sigma_{2}),
\end{eqnarray}
is satisfied for density matrices $\rho$, $\sigma_{1}$, and $\sigma_{2}$ numerically.
\section{Quantum speed limit time}

Now we are in a position to derive a new bound for QSL time by using the fidelity (6) as a distance measure introduced in the previous section. The absolute value for the time derivative of the fidelity $\mathcal{F}(\rho_{0},\rho_{t})$ is
\begin{eqnarray}
\begin{split}
\left|\frac{d\mathcal{F}}{dt}\right|
=\biggr|\frac{-Tr(\dot{\rho_{t}}\rho_{t})}{(Tr(\rho_{t}))^2}
\sqrt{\frac{1-Tr(\rho_{0}^2)}{Tr(\rho_{0}^2)}} \sqrt{\frac{Tr(\rho_{t}^2)}{1-Tr(\rho_{t}^2)}}Tr(\rho_{0}\rho_{t})
+
\\
\left(1+\sqrt{\frac{1-Tr(\rho_{0}^2)}{Tr(\rho_{0}^2)}}
\sqrt{\frac{1-Tr(\rho_{t}^2)}{Tr(\rho_{t}^2)}}\right)Tr(\rho_{0}\dot{\rho_{t}})\biggr|,
\end{split}
\end{eqnarray}
where by using triangle inequality it becomes as
\begin{eqnarray}
\begin{aligned}
\left|\frac{d\mathcal{F}}{dt}\right|
\leq
\sqrt{\frac{1-Tr(\rho_{0}^2)}{Tr(\rho_{0}^2)}}\sqrt{\frac{Tr(\rho_{t}^2)}{1-Tr(\rho_{t}^2)}}\left|\frac{Tr(\dot{\rho_{t}}\rho_{t})Tr(\rho_{0}\rho_{t})}{(Tr(\rho_{t}))^2}\right|
+
\\
\left(1+\sqrt{\frac{1-Tr(\rho_{0}^2)}{Tr(\rho_{0}^2)}}
\sqrt{\frac{1-Tr(\rho_{t}^2)}{Tr(\rho_{t}^2)}}\right)\left|Tr(\rho_{0}\dot{\rho_{t}})\right|.
\end{aligned}
\end{eqnarray}
By considering the Cauchy-Schwarz inequality in the second term of Eq. (23), we get
\begin{eqnarray}
\begin{split}
\biggr|\frac{d\mathcal{F}}{dt}\biggr|\leq\sqrt{\frac{1-Tr(\rho_{0}^2)}{Tr(\rho_{0}^2)}}\sqrt{\frac{Tr(\rho_{t}^2)}{1-Tr(\rho_{t}^2)}}\biggr|\frac{Tr(\dot{\rho_{t}}\rho_{t})Tr(\rho_{0}\rho_{t})}{(Tr(\rho_{t}))^2}\biggr|
+
\\
\sqrt{Tr(\rho_{0}^2)Tr(\dot{\rho_{t}}^2)}+
\sqrt{\frac{1-Tr(\rho_{t}^2)}{Tr(\rho_{t}^2)}}\sqrt{1-Tr(\rho_{0}^2)}\sqrt{Tr(\dot{\rho_{t}}^2)}.
\end{split}
\end{eqnarray}
Integration of Eq. (24) over deriving time $\tau$, gives the following inequality for the QSL time bound as follows

 \begin{eqnarray}
 \tau\geq\frac{|1-\mathcal{F_{\tau}}|}{X_{\tau}},
 \end{eqnarray}
 where $\mathcal{F_{\tau}}:=\mathcal{F}(\rho_{0},\rho_{\tau})$ is the target value of the fidelity at time $\tau$, and $X_{\tau}$ is defined as
 \begin{eqnarray}
 \begin{split}
 X_{\tau}:=\frac{1}{\tau}\int_{0}^{\tau}\biggr(\sqrt{\frac{1-Tr(\rho_{0}^2)}{Tr(\rho_{0}^2)}}\sqrt{\frac{Tr(\rho_{t}^2)}{1-Tr(\rho_{t}^2)}}\biggr|\frac{Tr(\dot{\rho_{t}}\rho_{t})Tr(\rho_{0}\rho_{t})}{(Tr(\rho_{t}))^2}\biggr|
+
\\
\sqrt{Tr(\rho_{0}^2)Tr(\dot{\rho_{t}}^2)}+
\sqrt{\frac{1-Tr(\rho_{t}^2)}{Tr(\rho_{t}^2)}}\sqrt{1-Tr(\rho_{0}^2)}\sqrt{Tr(\dot{\rho_{t}}^2)}\biggr)dt.
\end{split}
\end{eqnarray}
Eq. (25) provides an expression for lower bound of QSL time and can be used to consider for either Markovian or non-Markovian dynamics. It is interesting to note that in the case of initial pure states, we have $Tr(\rho_{0}^2)=1$, therefore Eq. (26) turns into
\begin{eqnarray}
X_{\tau}=\frac{1}{\tau}\int_{0}^{\tau}\sqrt{Tr(\dot{\rho_{t}}^2)}dt,
\end{eqnarray}
and the target fidelity becomes as
\begin{eqnarray}
\mathcal{F_{\tau}}=Tr(\rho_{0}\rho_{\tau}).
\end{eqnarray}
Substituting Eq. (27) and  Eq. (28) into Eq. (25), yields
\begin{eqnarray}
\tau\geq\frac{\tau|1-Tr(\rho_{0}\rho_{\tau})|}{\int_{0}^{\tau}\sqrt{Tr(\dot{\rho_{t}}^2)}dt},
\end{eqnarray}
which is the well-known Mandelstamm-Tamm type bound for nonunitary dynamics in the case of initial pure states, the case which was obtained initially in $\cite{Deffner}$ by using the Bures angle as a metric. In the next section, we examine our bound (25) by a concrete open quantum system as a physical model in which the initial state of the system is generally mixed.

\section{Physical model}

To investigate the performance of the bound (25) for QSL time, we consider a two-level quantum system which resonantly coupled to a leaky vacuum reservoir. The whole Hamiltonian of the system and the reservoir can be written as
\begin{eqnarray}
H=\frac{1}{2}\hbar\omega_{0}\sigma_{z}+\sum_{k}\hbar\omega_{k}a_{k}^\dag a_{k}+\sum_{k}\hbar(g_{k}a_{k}\sigma_{+}+g_{k}^*a_{k}^\dag \sigma_{-}),
\end{eqnarray}
where, $\sigma_{z}$ is the Pauli matrix and $\sigma_{+}$ ($\sigma_{-}$) is the Pauli raising (lowering) operator for the atom with transition frequency $\omega_{0}$. $a_{k}$ ($a_{k}^\dag$) is the annihilation (creation) operator for the $k$th field mode with frequency $\omega_{k}$ and $g_{k}$ is the coupling constant between the $k$th field mode and the system.
The dynamics of the system can be described by
\begin{eqnarray}
L_{t}(\rho_{t})=\gamma_{t}(\sigma_{-}\rho_{t}\sigma_{+}-\frac{1}{2}\{\sigma_{+}\sigma_{-},\rho_{t}\}).
\end{eqnarray}
The spectral density of the reservoir is assumed to have the Lorentzian form
\begin{eqnarray}
J(\omega)=\frac{1}{2\pi}\frac{\gamma_{0}\lambda^{2}}{(\omega_{0}-\omega)^{2}+\lambda^{2}},
\end{eqnarray}
where $\gamma_{0}$ is the coupling strength and $\lambda$ is the width of the Lorentzian function. The density matrix of the system at time $t$ can be obtained analytically $\cite{Breuer}$ as

\begin{eqnarray}
\rho(t)=\begin{pmatrix}
\rho_{11}(0)|G(t)|^2&\rho_{10}(0)G(t)\\
\rho_{01}(0)G(t)^*&1-\rho_{11}(0)|G(t)|^2\\
\end{pmatrix},
\end{eqnarray}
where the function $G(t)$ is define as the solution of the the integro-differential equation
\begin{eqnarray}
\frac{d}{dt}G(t)=-\int_{0}^{t} dt_{1}f(t-t_{1})G(t_{1}),
\end{eqnarray}
with the initial condition $G(0)=1$, and the correlation kernel $f(t-t_{1})$ related to the spectral density of the  reservoir as
 \begin{eqnarray}
f(t-t_{1})=\int d \omega J(\omega)e^{i(\omega_{0}-\omega_{k})(t-t_{1})}.
\end{eqnarray}
Using the Laplace transformation and its inverse, $G(t)$ can be given by
 \begin{eqnarray}
G(t)=e^{-\lambda t/2} [\cosh(\frac{dt}{2})+\frac{\lambda}{d} \sinh(\frac{dt}{2})],
\end{eqnarray}
with $d=\sqrt{\lambda^{2}-2\gamma_{0}\lambda}$. Also, the time dependent decay rate in Eq. (31) is given by $\gamma_{t}=-Im(\frac{\dot{G(t)}}{G(t)})$. The dynamics is Markovian in the weak-coupling regime $\gamma_{0}<\lambda/2$ and becomes non-Markovian for strong coupling $\gamma_{0}>\lambda/2$.
 In this work, we consider a mixed initial state of Werner-type
\begin{eqnarray}
 \rho(0)=\frac{1-r}{2}I+r|\psi\rangle\langle\psi|,
\end{eqnarray}
where $I$ is a $2\times2$ identity matrix, $0\leq r\leq1$ and $|\psi\rangle=(|1\rangle+|0\rangle)/\sqrt{2}$.

In Fig. 1 and Fig. 2, we present the QSL time bounds as a function of the coupling strength $\gamma_{0}$ for the initial states of Eq. (37) with different mixed coefficients $r$. Fig. 1a represent the QSL time bounds in Eq. (25) with parameter $\lambda=1$ and the deriving time $\tau=1$. We can see that the larger value of $r$ which is correspond to the higher purity, induce higher QSL time bound. As the non-Markovianity behavior grows up in term of $\gamma_{0}$, the lower bound decreases with respect to the mixedness of the initial state, i.e. the speed of evolution for the initial mixed states grows up. Fig. 1b sketches the obtained bound from the previous work $\cite{Zhe Sun}$ with the same condition of Fig. 1a, which is derived from exploiting $F_{1}$, i.e. Eq. (2), as a distance measure. Obviously, the QSL bound obtained in this paper is more tighter than the derived bound in the previous work $\cite{Zhe Sun}$ for both pure and mixed initial states.

Also, we reexamine our bound with $\lambda=20$ and $\tau=1$, and compare it with bound of $\cite{Zhe Sun}$, as depicted in Fig. 2a and Fig. 2b. Interestingly, for a given $r$, it is observed that the new bound not only is again tighter than the bound of $\cite{Zhe Sun}$ but also it becomes more sharper than the case brought in Fig. 2b. On the other hand, the sharply decrement of bound (25), when the environment enters to the non-Markovian regime is more apparent. Therefore, it can be treated as a better witness of non-Markovinity in this way.

\section{Conclusions}
We have introduced an alternative alternative fidelity which satisfies Jozsa's four axioms. We have shown that this fidelity is zero for any two orthogonal density matrices. Then by applying this fidelity as a distance measure between initial and time evolved final states of a quantum system, we have derived an improved bound for QSL time in open quantum systems. We have demonstrated that the improved bound leads to the well-known Mandelstamm-Tamm type bound for nonunitary dynamics in the case of initial pure states. Also, we have shown that in the case of initial mixed states, the bound (25) is tighter than the obtained bounds in the previous works. And finally, we have demonstrated that the improved bound decreases sharply in the non-Markovian regime.

\newpage
Fig. 1. QSL time, $\tau_{QSL}$, as a function of the coupling strength $\gamma_{0}$ for initial states (37) with different mixed coefficients $r=0.1, 0.5, 0.9, 1$. (a) The bounds derived from Eq. (25) and (b) the bounds obtained from [6]. $\lambda=1$, $\omega_{0}=1$, and  $\tau=1$.
\begin{figure}
\qquad\qquad\qquad\qquad a \qquad\qquad\qquad\qquad\qquad\qquad\qquad\qquad\qquad\qquad b \\
        {
        \includegraphics[width=3.3in]{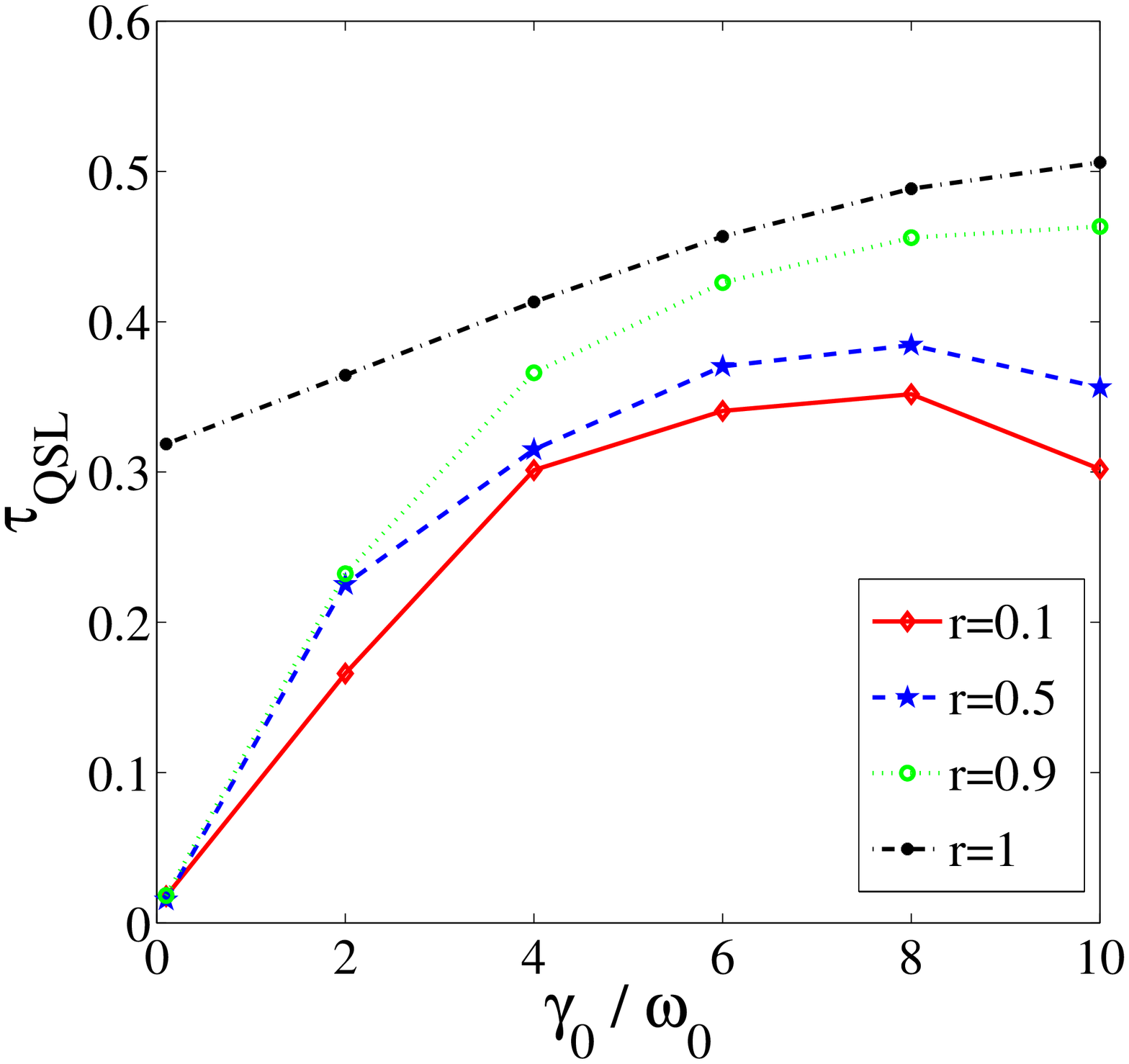}
        \label{fig:first_sub}
    }{
        \includegraphics[width=3.3in]{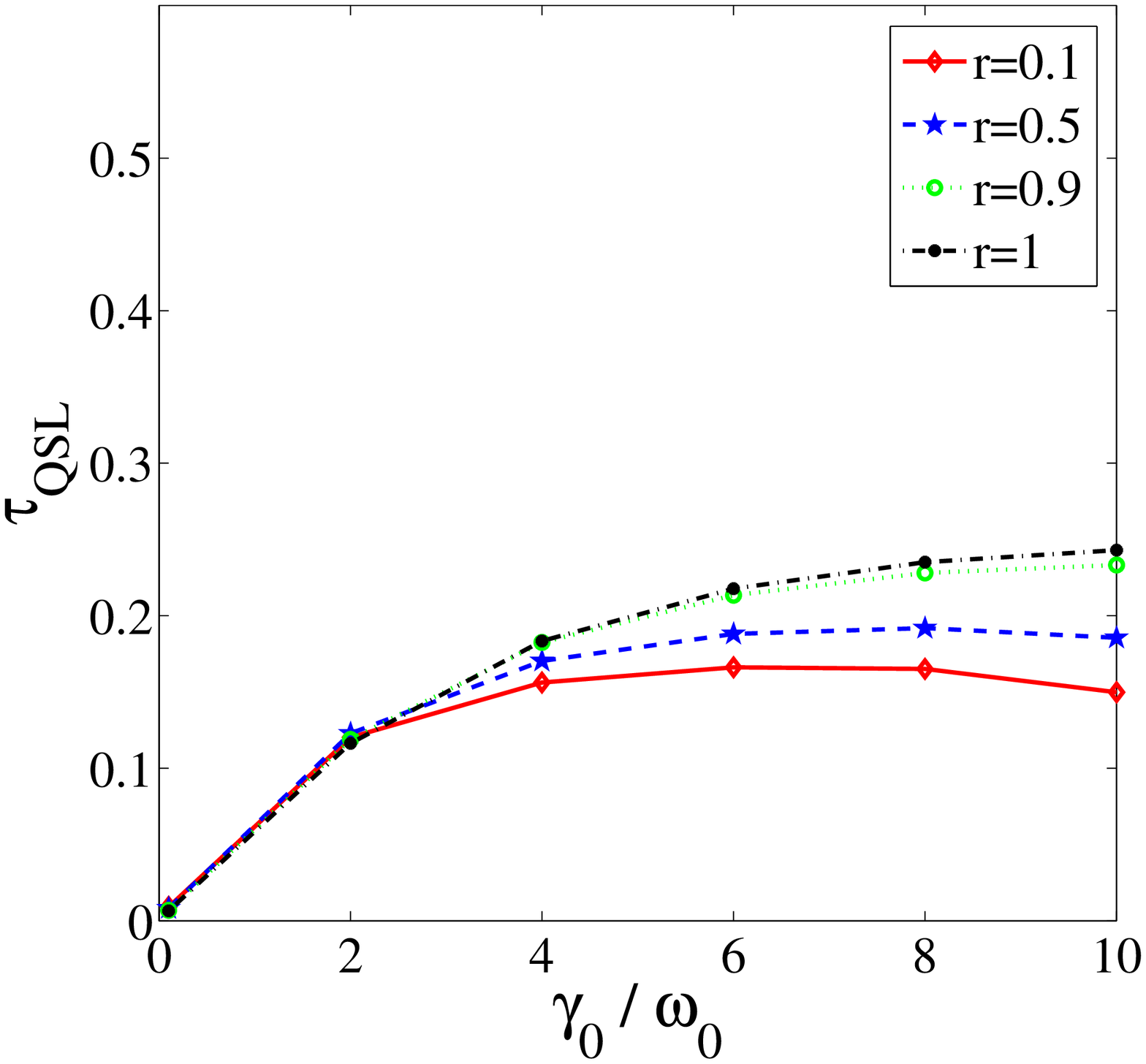}
        \label{fig:second_sub}
    }\\
        \caption{}
\end{figure}

\newpage
Fig. 2. QSL time, $\tau_{QSL}$, as a function of the coupling strength $\gamma_{0}$ for initial states (37) with different mixed coefficients $r=0.1, 0.5, 0.9, 1$. (a) The bounds derived from Eq. (25) and (b) the bounds obtained from [6]. $\lambda=20$, $\omega_{0}=1$, and  $\tau=1$.
\begin{figure}
\qquad\qquad\qquad\qquad a \qquad\qquad\qquad\qquad\qquad\qquad\qquad\qquad\qquad\qquad b \\
        {
        \includegraphics[width=3.3in]{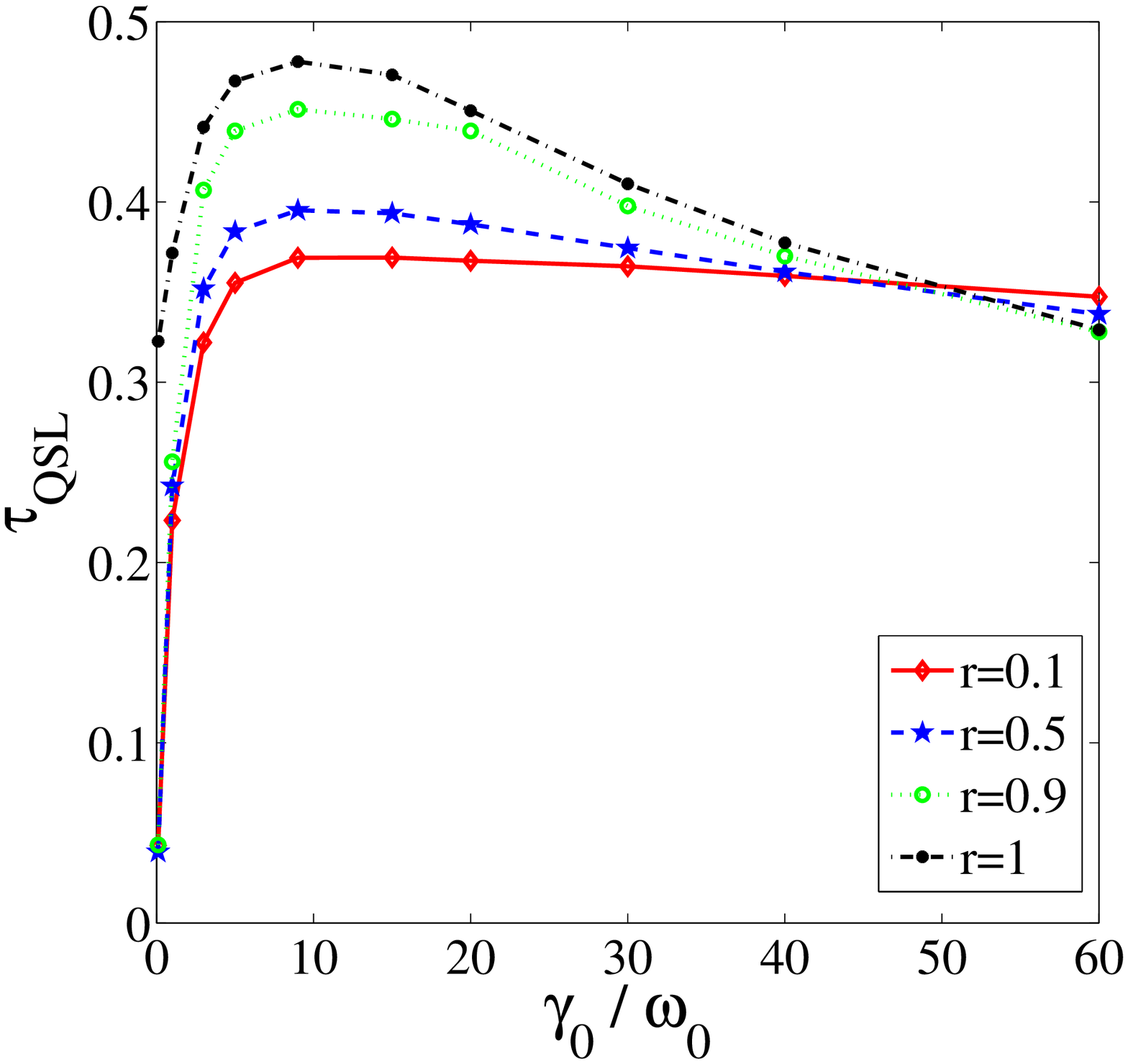}
        \label{fig:first_sub}
    }{
        \includegraphics[width=3.3in]{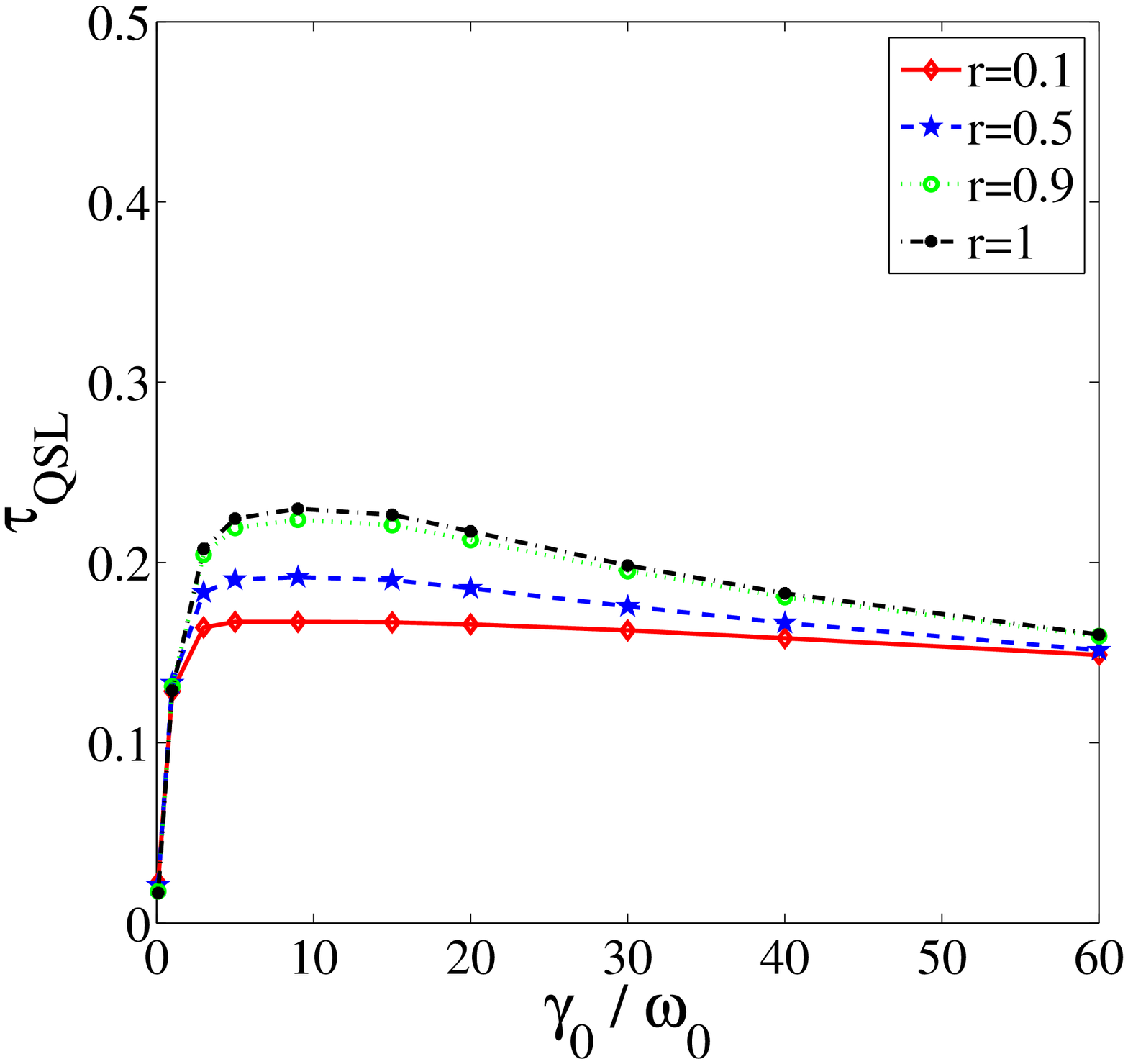}
        \label{fig:second_sub}
    }\\
        \caption{}
\end{figure}

\end{document}